\documentclass{article}
\usepackage{graphicx}
\begin{document}
\title{{\bf Einstein-Cartan cosmology and the $S_{8}$ problem}}
\author{{\bf Davor Palle} \\
ul. Ljudevita Gaja 35, 10000 Zagreb, Croatia \\
email: davor.palle@gmail.com}
\maketitle
\begin{abstract}
{
The measurements of cluster abundances, gravitational lensings, redshift space
distortions and peculiar velocities at lower redshifts point out to much smaller
$\sigma_{8}$ than its value deduced from the measurements of the CMB
fluctuations assuming the standard LCDM cosmology.
High redshift measurements of ALMA and JWST imply even more striking problems for LCDM.
We examine and compare the
$\sigma_{8}$ redshift dependence calculated within the gauge invariant formalism.
Because the CMB fluctuations comprise a cosmological data from the recombination
era to the present, the $S_{8}$ problem of the LCDM cosmology is not a surprise from
the standpoint of the Einstein-Cartan cosmology because it predicts much larger mass density
and $\sigma_{8}(z)$ than the LCDM model at high redshifts.
}
\end{abstract}

\section{Introduction and motivation}
The theorists are forced to invent new models or theories when the widely accepted
theories are confronted with the experimental and observational anomalies or
theoretical inconsistencies.
The Standard Model (SM) in particle physics and the standard LCDM (Lambda Cold Dark
Matter) model of the Universe are facing the common two theoretical problems:
(1) zero-distance singularity and (2) potential violation of causality.

The ultraviolet divergences in particle physics could be resolved by the Wick's theorem
and the hypothesis of the noncontractible space
without introducing the Higgs mechanism \cite{Palle1}, while the zero-distance singularity
in cosmology and gravity could be resolved within the Einstein-Cartan (EC) theory \cite{Trautman}
of Kibble and Sciama based on the Riemann-Cartan geometry. The potential violation of
causality and unitarity in particle physics, known as a $SU(2)$ global anomaly, is circumvented
by the exact cancellation of the $SU(2)$ weak boson and fermion (lepton or quark) anomalous actions
\cite{Palle1} with implications on the corresponding mixing angles. The analogous potential
violation of causality in General Relativity (GR) \cite{Goedel} is possible to avoid within
the EC cosmology \cite{Obukhov,Palle2} with the rotating Universe and the appropriate choice of
the chirality of the vorticity of the Universe \cite{Palle2,Palle3}.

The general remark is that the profound equilibrium between the translational and rotational degrees
of freedom in cosmology and the boson and fermion degrees of freedom in particle physics
permits the mathematical and phenomenological consistent physical theories in Minkowski
and Riemann-Cartan spacetimes. There is no need for the inflaton scalar in cosmology or
the Higgs scalar in particle physics. The models with scalars do not solve the mass density
problem in cosmology or the masses of elementary particles. Only new principle of the
noncontractible space guides to properly defined EC equations \cite{Obukhov,Palle2,Palle3,Hehl} and the nonsingular
Dyson-Schwinger and Bethe-Salpeter equations in particle physics derived from the generalized
Wick's theorem \cite{Palle1}.

In the next chapter we describe in more detail our setup in particle physics and cosmology.
The third chapter describes a derivation of equations necessary to resolve the $S_{8}$ problem.
We discuss and comment the numerical results in the final section.

\section{Resume of the theoretical setup in Minkowski and Riemann-Cartan spacetimes}
The SM (one Higgs doublet) masses of elementary particles are unknown since they
are defined by the free parameters: Yukawa couplings and the Higgs potential.
However, there is one exception: the SM predicts that Dirac neutrino masses vanish
$m_{\nu}=m_{\bar{\nu}}=0$. On the contrary, the neutrinos in our theory \cite{Palle1} are Majorana
massive particles. Light Majorana neutrinos have large number densities and interact with torsion
because they have masses \cite{Hehl} (unlike massless photons) and are responsible to
generate primordial vorticity \cite{Palle3} as a hot dark matter; heavy Majorana neutrinos
are cosmologically stable \cite{Palle4} cold dark matter particles. They can produce
cosmological excess of the lepton and baryon numbers \cite{Fukugita} but with the scalar
doublet which contains three unphysicsl Nambu-Goldstone scalars and
one unphysical $\zeta$ particle \cite{Palle5}. The masses of heavy and light neutrinos
match the estimate of the required cosmological baryon to photon number density
ratio $\eta$ \cite{Fukugita}.
It follows that our theory of noncontractible space in particle physics has three necessary
ingredients for cosmology: (1) light massive neutrinos, (2) heavy neutrinos as a cold dark
matter particles and (3) violation of baryon and lepton numbers. Because it contains a new
fundamental constant of Nature, the UV cutoff $\Lambda_{UV} = \frac{\hbar}{c d}$ that
defines the minimal distance $d$, we have to examine its universality within the
nonsingular EC cosmology \cite{Trautman}.

The minimal distance d in cosmology is fixed by the extremum condition of the EC effective
matter and radiation density as the zero'th order in perturbation cosmic parameters
$d\simeq 2 R_{min}$ (R is the cosmological scale factor).
One can neglect very small contributions from vorticity and acceleration.
If we account for the most precise evaluation of the $M_{W}$ and the UV cut-off $\Lambda_{UV}$
that defines the minimal distance
within our theory \cite{Palle5,New1}, it is possible to estimate the number
density of the primordial particle with spin $\hbar/2$ \cite{Trautman,Palle2} (subscript 0 denotes
present values):
\begin{eqnarray}
&&\dot{R}(R_{min})=0\ \Rightarrow \kappa \rho_{\gamma}(R_{min})=\kappa^{2}\frac{\hbar^{2}}{4}n^{2}(R_{min}), \\
&&S=\frac{\hbar}{2}n=scalar\ spin\ density\ of\ primordial\ particle,\ n=number\ density, \nonumber \\
&&Q=\kappa S=scalar\ torsion \nonumber \\
&&\Rightarrow n_{0}\simeq \frac{2 R_{min}}{R_{0}\hbar}(\frac{\rho_{\gamma,0}}{\kappa})^{1/2},\
d\simeq 2 R_{min},\ R_{0}\simeq c/H_{0},\ \kappa=8\pi G_{N}c^{-4}, \nonumber \\
&&\Lambda_{UV} = \frac{\hbar}{c d} =\frac{\pi}{\sqrt{6}}\frac{2}{g} M_{W},\
 e = g \sin \Theta_{W},\ \cos \Theta_{W} = \frac{M_{W}}{M_{Z}},\ \Omega_{CDM}(EC)\simeq 2, \nonumber \\
&& M_{W}=W\ weak\ boson\ mass,\ g=weak\ coupling,\ \Theta_{W}=Weinberg\ angle
\nonumber \\
&&\Rightarrow n_{0}\simeq {\cal O}(10^{-10} cm^{-3})
\Rightarrow m_{CDM}=\rho_{CDM,0}/n_{0}\simeq {\cal O}(100 TeV). \nonumber
\end{eqnarray}

The resulting mass of the primordial particle is close to the expectations for the heavy Majorana
neutrino masses and compatible with the unitarity bound for CDM particles \cite{Griest}.
It is worth to mention the HESS galactic center gamma source J1745-290 as a viable candidate for
the annihilation of the CDM particle because it is on the right place, with no time variability but
with the characteristic CDM annihilation spectrum \cite{HESS,Cembranos}.

The Minkowski spacetime dimension was crucial for our conformal $SU(3)$ unification \cite{Palle1}.
It is well known that the conformal Weyl tensor does not vanish for spacetimes $n \geq 4$.
We use Penrose's conformal technique \cite{Penrose} to fix the mass density near the
spacelike infinity in the EC cosmology ($\Omega_{m}=2$) \cite{Palle2}. There is no physical process that can
initiate substantial curvature on the hypersurface orthogonal to the expansion. Small ammount of vorticity and
acceleration are necessary \cite{Palle3,Palle6} general relativistic invariants to generate torsion
of spacetime. The nonrelativistic limes of torsion is the angular momentum of the Universe developed
along the evolution and growth of galaxies and galaxy clusters because of the predominanace
of the right-handed structures \cite{Palle3}. The torsion acts opposite to the matter density
and is hidden to our observations obeying the same evolution law as that of the nonrelativistic matter mass
density \cite{Palle6}.

While the SU(2) global anomaly cancellation in particle physics leads to the relation between
the fermion mixing and electroweak boson mixing angles \cite{Palle1,Palle5}, the left handed
chirality of the weak interactions induces the right handed vorticity of the Universe within
the EC cosmology
$\chi(weak\ interactions)+\chi(vorticity\ of\ the\ Universe)=-1+1=0$ if the right handed coordinate
system is used. If we use the left handed coordinate system the cancellation of the
chiralities is preserved $\chi(weak\ interactions)+\chi(vorticity\ of\ the\ Universe)=+1-1=0$,
albeit with the opposite chiralities \cite{Palle3}.

The angular correlations of the pulsar timing arrays
allow to estimate the present vorticity of the Universe to be $\frac{\omega_{0}}{H_{0}}=
{\cal O}(10^{-5})$ inducing the rotation of the CMB polarization vector compatible with measurements \cite{Palle7}.

We demonstrate that the spin densities through torsion activate the primordial density contrast
\cite{Palle8} in the EC cosmology without the need to introduce some scalar inflaton as in
the LCDM.

Knowing the particle content of the Universe and the structure of the Einstein-Cartan cosmology
that respects and relates translational and rotational degrees of freedom of matter, we derive
necessary equations to calculate $\sigma_{8}(z)$ observable in the next section.

\section{Derivation of perturbation equations}
We start with the EC field equations for the Weysssenhoff fluid to the leading order without
small contributions of vorticity or acceleration \cite{Palle6,Palle9,Palle10,Palle11,New2}:
\begin{eqnarray}
&&3\frac{\dot{R}^{2}}{R^{2}} = \kappa (\rho + \Lambda)-Q^{2},\hspace{50mm} \nonumber \\
&&2\frac{\ddot{R}}{R}+\frac{\dot{R}^{2}}{R^{2}}= \kappa (\Lambda-p)+Q^{2},
\ \kappa=8\pi G_{N}c^{-4} \nonumber \\
&&Q^{2}=\frac{1}{2}Q_{\mu\nu}Q^{\mu\nu},\ Q^{\mu}_{\ \nu\rho}=u^{\mu}Q_{\nu\rho},\
u^{\mu}Q_{\mu\rho}=0, \nonumber \\
&&Q^{\mu}_{\ ab}+2v^{\mu}_{[a}Q_{b]}=\kappa S^{\mu}_{\ ab},\ Q_{a}=v^{\mu}_{a}Q_{\mu},\
Q_{\mu}=Q^{\nu}_{\ \mu\nu}, \\
&&\rho=\rho_{(c)}+\rho_{(b)}+\rho_{(\gamma)}+\rho_{(\nu)},\
p=p_{(c)}+p_{(b)}+p_{(\gamma)}+p_{(\nu)}, \nonumber \\
&&g^{\mu\nu} = v^{\mu}_{a}v^{\nu}_{b}\eta^{a b},\
\eta_{a b}=diag(+1,-1,-1,-1),\
 \mu,\nu=0,1,2,3, \ a,b=\hat{0},\hat{1},\hat{2},\hat{3}, \nonumber \\
&& Q^{\mu}_{.\nu\lambda}=torsion\ tensor,\ S^{\mu}_{.\nu\lambda}=spin-angular\ momentum\ tensor, \nonumber \\
&&u^{\mu}=velocity\ four\ vector,\ v^{\mu}_{a}=Vierbein\ vectors, \nonumber \\
&&\rho=energy\ density,\ p=pressure\ density,\ \Lambda=cosmological\ constant. \nonumber
\end{eqnarray}

Although we neglect vorticity and acceleration to the leading order of EC equations, we
preserve acceleration in linear perturbations.
Vorticity ($\omega=m=0$) is ignored in this work because it is irelevant
for the isotropic $\sigma_{8}(z)$ observable, but we retain in our derivations $\tilde{\omega}_{\mu\nu}$ since it
contains torsion.

It is assumed that we include matter density and pressure of the CDM,
massless neutrinos, baryons and photons (denoted in brackets respectively as (c),($\nu$),(b),($\gamma$)).
We ignore light neutrino masses because their influence on $\sigma_{8}(z)$ is negligible but not to
the CMB anisotropies which are very small but precise cosmological observables.
The metric includes expansion, acceleration and vorticity:
\begin{eqnarray}
d s^{2}=d t^{2}-R^{2}(t)[d x^{2}+(1-\lambda^{2}(t))e^{2mx}d y^{2}]
-R^{2}(t)d z^{2}-2 R(t)\lambda(t)e^{mx}d y dt,
\end{eqnarray}
\begin{eqnarray*}
\frac{\dot{R}}{R}=H=\frac{\Theta}{3},\ a^{\mu}a_{\mu}=-(\lambda H+\dot{\lambda})^{2},
\ \omega^{2}=\frac{1}{2}\omega_{\mu\nu}\omega^{\mu\nu}=(\frac{\lambda m}{2 R})^{2},\
 m=const.
\end{eqnarray*}

The standard Ehlers-decomposition of the velocity-gradient can be written as

\begin{eqnarray}
&&\tilde{\nabla}_{\mu}u_{\nu}=\tilde{\omega}_{\nu\mu}
+\sigma_{\mu\nu}+\frac{1}{3}\Theta h_{\mu\nu}+u_{\mu}a_{\nu}, \\
&& u^{\mu}u_{\mu}=1,\ h_{\mu\nu}=g_{\mu\nu}-u_{\mu}u_{\nu},
\ a_{\mu}=u^{\nu}\tilde{\nabla}_{\nu}u_{\mu},
\ \Theta=\tilde{\nabla}_{\nu}u^{\nu}, \nonumber \\
&&\tilde{\omega}_{\mu\nu}=h^{\alpha}_{\mu}h^{\beta}_{\nu}
\tilde{\nabla}_{[\beta}u_{\alpha ]},\
\sigma_{\mu\nu}=h^{\alpha}_{\mu}h^{\beta}_{\nu}
\tilde{\nabla}_{(\alpha}u_{\beta )}-\frac{1}{3}\Theta h_{\mu\nu}, \nonumber
\end{eqnarray}
with definitions:
\begin{eqnarray}
&&\tilde{\Gamma}^{\alpha}_{\beta\mu}\equiv\Gamma^{\alpha}_{\beta\mu}
+Q^{\alpha}_{. \beta\mu}+Q_{\beta\mu .}^{\ .  . \ \alpha}+
Q_{\mu\beta .}^{\ .  .  \ \alpha},\ \Gamma^{\alpha}_{\beta\mu}=Christoffel\ symbol, \nonumber \\
&&\tilde{\nabla}_{\alpha}u_{\beta}\equiv\partial_{\alpha}u_{\beta}
-\tilde{\Gamma}^{\nu}_{\beta\alpha}u_{\nu},\
\nabla_{\alpha}u_{\beta}\equiv\partial_{\alpha}u_{\beta}
-\Gamma^{\nu}_{\beta\alpha}u_{\nu}. \nonumber
\end{eqnarray}

We use the Bianchi and Ricci identities (note the wrong sign in front of the Mathisson-Papapetrou term
of the Bianchi identity in \cite{Obukhov}):
\begin{eqnarray}
&&(\tilde{\nabla}_{\nu}-2 Q_{\nu})T^{\nu}_{. \mu}+2Q^{\alpha}_{.\mu\beta}T^{\beta}_{. \alpha}
-S^{\nu}_{.\alpha\beta}\tilde{R}^{\alpha\beta}_{. . \mu\nu}=0, \\
&&\tilde{R}^{\lambda}_{. \sigma\mu\nu} \equiv
\partial_{\mu} \tilde{\Gamma}^{\lambda}_{\sigma\nu}
-\partial_{\nu} \tilde{\Gamma}^{\lambda}_{\sigma\mu}
+\tilde{\Gamma}^{\lambda}_{\beta\mu}\tilde{\Gamma}^{\beta}_{\sigma\nu}
-\tilde{\Gamma}^{\lambda}_{\beta\nu}\tilde{\Gamma}^{\beta}_{\sigma\mu}, \nonumber \\
&&(\tilde{\nabla}_{\mu}\tilde{\nabla}_{\nu}
-\tilde{\nabla}_{\nu}\tilde{\nabla}_{\mu})u_{\lambda}=
-\tilde{R}^{\sigma}_{. \lambda\mu\nu}u_{\sigma}
-2 Q^{\sigma}_{. \nu\mu}\tilde{\nabla}_{\sigma}u_{\lambda}.
\end{eqnarray}

The following identity derived from the Ricci identity is employed (Frenkel's condition
is included $Q^{\mu\nu}u_{\mu}=0$):
\begin{eqnarray}
&&^{(3)}\tilde{\nabla}_{\mu}(\dot{f})-h^{\nu}_{\mu}(^{(3)}\tilde{\nabla}_{\nu}
f)^{.}
=a_{\mu}\dot{f}+(\tilde{\omega}^{\lambda}_{. \mu}+
\sigma_{\mu .}^{\ \lambda}+\frac{1}{3}\Theta h_{\mu}^{\lambda})
^{(3)}\tilde{\nabla}_{\lambda}f, \\
&&^{(3)}\tilde{\nabla}_{\mu}f \equiv h^{\nu}_{\mu}\tilde{\nabla}_{\nu}f,\
^{(3)}\tilde{\nabla}_{\mu}X_{\lambda} \equiv h^{\nu}_{\mu}h^{\kappa}_{\lambda}\tilde{\nabla}_{\nu}X_{\kappa},
\ \dot{X_{\mu}}\equiv u^{\nu}\tilde{\nabla}_{\nu}X_{\mu}, \nonumber \\
&& f=arbitrary\ scalar,\ X_{\mu}=arbitrary\ vector. \nonumber
\end{eqnarray}

We show in the Appendix A that the standard decomposition of the conformal tensor is not valid
if $\frac{d \lambda}{d t}\omega \neq 0$. Thus, we do not refer to this decomposition in our derivation
of equations for perturbed quantities.

It is necessary to fix the form of the gauge invariant variables introduced by Ellis et al.
\cite{Ellis1,Ellis2,Challinor}.
This problem we address in the Appendix B.

All the equations are valid only to linear order in perturbed quantities.
The equation for the gauge invariant density contrast can now be derived from the Bianchi identity and
the identity Eq.(7) ((i)=(b) or (c), $w_{i}=\frac{p_{i}}{\rho_{i}}$):
\begin{eqnarray}
\dot{\cal D}_{(i)\mu}&=&-(1+w_{i})[Z_{\mu}+R ^{(3)}\tilde{\nabla}_{\mu}
(\tilde{\nabla}_{\lambda}v^{\lambda}_{(i)})]+R(1+w_{i})\Theta a_{\mu} \nonumber \\
&& -\tilde{\omega}^{\lambda}_{\ \mu}{\cal D}_{(i)\lambda}
-\rho^{-1}_{(i)}R\Theta ^{(3)}\tilde{\nabla}_{\mu}p_{(i)}+\Theta w_{i}{\cal D}_{(i)\mu}, \\
&&{\cal D}_{(i)\mu}\equiv \rho^{-1}_{(i)}R^{(3)}\tilde{\nabla}_{\mu}\rho_{(i)}=gauge\ invariant\
density\ contrast\ vector, \nonumber \\
&& Z_{(i)\mu}\equiv R^{(3)}\tilde{\nabla}_{\mu}\Theta=gauge\ invariant\ perturbed\ expansion\ vector.
\nonumber
\end{eqnarray}

Note that for any vector $X_{\mu}=(0,X_{i})$ the following relation holds to linear order
irrespective of the scalar, vector or tensor spatial perturbations \cite{Kodama}:
\begin{eqnarray}
\dot{X_{\mu}}\equiv u^{\nu}\tilde{\nabla}_{\nu}X_{\mu}=(0,\frac{\partial X_{1}}{\partial t}-H X_{1}
+Q X_{2},\frac{\partial X_{2}}{\partial t}-H X_{2}-Q X_{1},\frac{\partial X_{3}}{\partial t}-H X_{3}).
\end{eqnarray}

From the Eq.(7), EC field equations and Ricci identities, we derive the evolution equation
for $Z_{\mu}$:
\begin{eqnarray}
&&\dot{Z}_{\mu}=4 Q \Psi_{\mu}-\frac{2}{3}\Theta Z_{\mu}-\frac{1}{2}\sum_{i}\kappa \rho_{i}
(1+3w_{i}){\cal D}_{(i)\mu}-u_{\mu}Z_{\nu}a^{\nu} \nonumber \\
&&\ \ \ \ \ \ -\dot{\Theta}a_{\mu}R
-\tilde{\omega}^{\lambda}_{\ \mu}Z_{\lambda}, \\
&&\Psi_{\mu}\equiv R ^{(3)}\tilde{\nabla}_{\mu}Q,\ \dot{\Theta}=2 Q^{2}-\frac{1}{3}\Theta^{2}
+\kappa \Lambda -\frac{1}{2}\sum_{i}\kappa \rho_{i}(1+3w_{i}), \nonumber \\
&&Q\Psi_{\mu}=\frac{1}{2}\sum_{i}\kappa \rho_{i}{\cal D}_{(i)\mu}
-\frac{1}{3}\Theta Z_{\mu}, \ for\ Q\neq 0. \nonumber
\end{eqnarray}

We derive photon field perturbations from the geodesic equations (remembering that massless photons do not
interact with torsion \cite{Hehl}) \cite{Ellis3,Ma,Challinor}:
\begin{eqnarray}
&&{\cal L}f(x^{\mu},p^{\mu})\equiv \frac{d x^{\mu}}{d\sigma}\frac{\partial f}{\partial x^{\mu}}
+\frac{d p^{\mu}}{d\sigma}\frac{\partial f}{\partial p^{\mu}}=C_{f}, \\
&& p^{\mu}\equiv\frac{d x^{\mu}}{d \sigma},\ p^{\mu}\nabla_{\mu}p^{\nu}=0,\
\frac{d p^{\mu}}{d \sigma}+\Gamma^{\mu}_{\nu\lambda}p^{\nu}p^{\lambda}=0, \nonumber \\
&& \Rightarrow p^{\mu}\frac{\partial f}{\partial x^{\mu}}-\Gamma^{\mu}_{\nu\lambda}
p^{\nu}p^{\lambda} \frac{\partial f}{\partial p^{\mu}}=C_{f}, \nonumber
\end{eqnarray}
\begin{eqnarray*}
p_{a}=E(u_{a}+e_{a}),\ p_{a}p^{a}=0,\ u_{a}u^{a}=1,\ e_{a}e^{a}=-1,\ e_{a}u^{a}=0, \\
p^{\mu}=v^{\mu}_{b}p^{b},\ \partial_{b}\equiv v^{\mu}_{b}\frac{\partial}
{\partial x^{\mu}},\ \Gamma^{a}_{bc}\equiv \Gamma^{\mu}_{\nu\lambda}
v^{a}_{\mu}v^{\nu}_{b}v^{\lambda}_{c} \\
\Rightarrow p^{b}\partial_{b}f-\Gamma^{a}_{bc}p^{b}p^{c}(u_{a}\frac{\partial f}{\partial E}
+E^{-1}\frac{\partial f}{\partial e^{a}})=C_{f}.
\end{eqnarray*}

With a definitions:
\begin{eqnarray*}
\rho^{(\gamma)}\equiv \int dEd\Omega E^{3}f(E,e^{b}),\
q^{(\gamma)}_{a}\equiv \int dEd\Omega E^{3}f(E,e^{b})e_{a},
\end{eqnarray*}

we get the equations in the linear approximation ($\sigma_{T}$=Thomson cross section):
\begin{eqnarray}
&&u^{b}\partial_{b}\rho^{(\gamma)}+\frac{4}{3}\Theta\rho^{(\gamma)}
+h^{ab}\partial_{b}q_{a}^{(\gamma)}=0 \nonumber \\
&&\Rightarrow \frac{\partial{\cal D}_{(\gamma)\mu}}{\partial t}
=H {\cal D}_{(\gamma)\mu}-\frac{4}{3}Z_{\mu}+\frac{4}{3}\Theta R a_{\mu}
-R\rho_{(\gamma)}^{-1\ (3)}\nabla_{\mu}(\partial_{\nu}q_{(\gamma)}^{\nu}), \\
&&u^{b}\partial_{b}q^{c}_{(\gamma)}+\frac{2}{3}\Theta q^{c}_{(\gamma)}
-\frac{1}{3}h^{cd}\partial_{d}\rho_{(\gamma)}=n_{e}\sigma_{T}
(\frac{4}{3}\rho_{(\gamma)}v^{c}_{(b)}-q^{c}_{(\gamma)}).
\end{eqnarray}

It is left to derive evolution equations for CDM and baryon velocities defined by
$u_{\mu}^{(i)}=u_{\mu}+v_{\mu}^{(i)}$ \cite{Challinor}. To linear approximation
in $v^{(b)\mu}$ the Bianchi identity is:
\begin{eqnarray}
2\tilde{\nabla}_{\nu}[(\rho^{(b)}+p^{(b)})u^{(\nu}v^{(b)}_{\mu)}
+2v^{(b)(\nu}u^{\alpha)}\tilde{\nabla}_{\beta}(u^{\beta}S_{\alpha\mu}) \nonumber \\
+u^{\nu}u^{\alpha}\tilde{\nabla}_{\beta}(v^{(b)\beta}S_{\alpha\mu})]
-S_{\alpha\beta}\tilde{R}^{\alpha\beta}_{..\mu\nu}v^{(b)\nu}=0
\end{eqnarray}

The spacelike vector $v^{(b)}_{\mu}=(0,v^{(b)}_{1},v^{(b)}_{2},v^{(b)}_{3})$ of the baryon velocity
evolves according to ($w_{b}=0$):
\begin{eqnarray}
\frac{\partial v_{1}^{(b)}}{\partial t}&=&(G_{1}^{2}+G_{2}^{2})^{-1}[(G_{2}G_{3}-G_{1}G_{4})v_{2}^{(b)}
-(G_{1}G_{3}+G_{2}G_{4})v_{1}^{(b)}+G_{2}H_{2}^{(b)}-G_{1}H_{1}^{(b)}], \nonumber \\
\frac{\partial v_{2}^{(b)}}{\partial t}&=&(G_{1}^{2}+G_{2}^{2})^{-1}[(G_{1}G_{4}-G_{2}G_{3})v_{1}^{(b)}
-(G_{1}G_{3}+G_{2}G_{4})v_{2}^{(b)}-G_{1}H_{2}^{(b)}-G_{2}H_{1}^{(b)}], \nonumber \\
\frac{\partial v_{3}^{(b)}}{\partial t}&=&\Theta c_{s}^{2}v_{3}^{(b)}
-\rho_{(b)}^{-1}n_{e}\sigma_{T}(\frac{4}{3}\rho_{(\gamma)}v_{3}^{(b)}-q_{3}^{(\gamma)}),
 \\
G_{1}&=&\kappa\rho_{(b)}+2Q^{2},\ G_{2}=2\dot{Q}+4Q H,\ G_{3}=-c_{s}^{2}\Theta \kappa \rho_{(b)}
+2Q\dot{Q}, \nonumber \\
G_{4}&=&2 Q \kappa\rho_{(b)}+2(4 H\dot{Q}+\ddot{Q}+Q H^{2}+2 \frac{\ddot{R}}{R}Q),\
H^{(b)}_{1,2}=\kappa n_{e}\sigma_{T}(\frac{4}{3}\rho_{(\gamma)}v_{1,2}^{(b)}-q_{1,2}^{(\gamma)}).
\nonumber
\end{eqnarray}

The acceleration function $\lambda (t)$ has evolution defined by the Bianchi identity
(however we use only the first derivative to fix the evolution):
\begin{eqnarray}
\dot{\lambda}&=&[2Q^{2}-\kappa (p+\rho)]^{-1}\lambda[\kappa (p+\rho)H+\kappa \dot{p}
-2(Q^{2}H+Q\dot{Q})], \\
\ddot{\lambda}&=&-\frac{\dot{Q}}{Q}(\dot{\lambda}+\lambda H)-3\lambda H^{2}-4\dot{\lambda}H.
\nonumber
\end{eqnarray}

The Thomson scattering term with $n_{e}=x_{e}n_{H}$ is evaluated using RECFAST code \cite{Seager}
(modified for the EC model),
as well as the other quantities necessary to evaluate $c_{s}^{2}$ \cite{Ma}. We include into the code
the reionization model based on the Planck data.
The CDM velocity vanishes like in the synchronous gauge \cite{Ma}.

\section{Results and discussion}
We restrict perturbations to the scalar spatial ones \cite{Kodama} and decompose our gauge invariant
quantities to the time and spatial parts \cite{Kodama,Ma,Challinor} acknowledging constraint
equations(for hypersurface with a flat metric the harmonic function is $Y^{(k)}(\vec{x})=
e^{\imath \vec{k}\cdot \vec{x}}$):
\begin{eqnarray}
&&{\cal D}^{(i)}_{\mu}(t,\vec{x})=\sum_{k} k \bar{{\cal D}}^{(i)}_{\mu}(t,k)Y^{(k)}(\vec{x}),\
i=c,b,\gamma ,\
Z_{\mu}(t,\vec{x})=\sum_{k} \frac{k^{2}}{R} \bar{Z}_{\mu}(t,k)Y^{(k)}(\vec{x}), \nonumber \\
&& q_{\mu}^{(\gamma)}(t,\vec{x})=\rho_{(\gamma)}\sum_{k} \bar{q}_{\mu}^{(\gamma)}(t,k)Y^{(k)}(\vec{x}),\
v_{\mu}^{(b)}(t,\vec{x})=\sum_{k} \bar{v}_{\mu}^{(b)}(t,k)Y^{(k)}(\vec{x}).
\end{eqnarray}

Including the above decompositions into the evolution equations we finally arrive to
the equations suitable for numerical computations (proper time variable t is replaced
by $y=\ln R$; all four-vectors are spacelike, hence we put zero to the time components):
\begin{eqnarray}
&&\frac{d \bar{{\cal D}}_{i}^{(\gamma)}}{d y}=\bar{{\cal D}}_{i}^{(\gamma)}-\frac{4}{3}\frac{k}{R H}
\bar{Z}_{i}+4 \frac{R}{k}a_{i}-\frac{k}{3 H R}\bar{q}_{i}^{(\gamma)}, \nonumber \\
&&\frac{d \bar{{\cal D}}_{i}^{(b)}}{d y}=\bar{{\cal D}}_{i}^{(b)}-\frac{k}{R H}
\bar{Z}_{i}+3 \frac{R}{k}a_{i}-\frac{k}{3 H R}\bar{v}_{i}^{(b)}, \nonumber \\
&&\frac{d \bar{{\cal D}}_{i}^{(c)}}{d y}=\bar{{\cal D}}_{i}^{(c)}-\frac{k}{R H}
\bar{Z}_{i}+3 \frac{R}{k}a_{i}, \nonumber \\
&&\frac{d \bar{Z}_{i}}{d y}=-\frac{R}{2Hk}\sum_{j}(1+3w_{j})\kappa \rho_{(j)}\bar{{\cal D}}_{i}^{(j)}
-\dot{\Theta}\frac{R^{2}}{k^{2} H}a_{i},\ for\ Q=0, \nonumber \\
&&\frac{d \bar{Z}_{i}}{d y}=-4 \bar{Z}_{i}
-\frac{3R}{2Hk}\sum_{j}(1-w_{j})\kappa \rho_{(j)}\bar{{\cal D}}_{i}^{(j)}
-\dot{\Theta}\frac{R^{2}}{k^{2} H}a_{i},\ for\ Q\neq0, \nonumber \\
&&\frac{d \bar{q}_{i}^{(\gamma)}}{d y}=3\bar{q}_{i}^{(\gamma)}
+\frac{1}{3}\frac{k}{R H}\bar{{\cal D}}_{i}^{(\gamma)}
+\frac{n_{e}\sigma_{T}}{H}(\frac{4}{3}\bar{v}^{(b)}_{i}-\bar{q}_{i}^{(\gamma)}), \nonumber
\end{eqnarray}
\begin{eqnarray}
&&\frac{d \bar{v_{1}}^{(b)}}{d y}=H^{-1}(G_{1}^{2}+G_{2}^{2})^{-1}[(G_{2}G_{3}-G_{1}G_{4})\bar{v}_{2}^{(b)}
-(G_{1}G_{3}+G_{2}G_{4})\bar{v}_{1}^{(b)}+G_{2}\bar{H}_{2}^{(b)}-G_{1}\bar{H}_{1}^{(b)}], \nonumber \\
&&\frac{d \bar{v}_{2}^{(b)}}{d y}=H^{-1}(G_{1}^{2}+G_{2}^{2})^{-1}[(G_{1}G_{4}-G_{2}G_{3})\bar{v}_{1}^{(b)}
-(G_{1}G_{3}+G_{2}G_{4})\bar{v}_{2}^{(b)}-G_{1}\bar{H}_{2}^{(b)}-G_{2}\bar{H}_{1}^{(b)}], \nonumber \\
&&\frac{d \bar{v}_{3}^{(b)}}{d y}=H^{-1}\Theta c_{s}^{2}\bar{v}_{3}^{(b)}
-H^{-1}\rho_{(b)}^{-1}n_{e}\sigma_{T}\rho_{(\gamma)}(\frac{4}{3}\bar{v}_{3}^{(b)}-\bar{q}_{3}^{(\gamma)}), \\
&&\bar{H}^{(b)}_{1,2}=\kappa n_{e}\sigma_{T}\rho_{(\gamma)}(\frac{4}{3}\bar{v}_{1,2}^{(b)}-\bar{q}_{1,2}^{(\gamma)}).
\nonumber
\end{eqnarray}

The initial conditions are the standard ones derived from the set of equations valid at very high redshifts
\cite{Ma,Challinor}:

\begin{eqnarray}
&&\bar{\cal D}^{(c)}_{i}=\bar{\cal D}^{(b)}_{i}=\frac{3}{4}\bar{\cal D}^{(\gamma)}_{i}
=\frac{3}{4}\bar{\cal D}^{(\nu)}_{i},\ \bar{v}^{(b)}_{i}=\frac{3}{4}\bar{q}^{(\gamma)}_{i},
 \\
&& \bar{\cal D}^{(\gamma)}_{i}=D x^{2},\ \bar{q}^{(\gamma)}_{i}=\frac{D}{9}x^{3},\
\bar{Z}_{i}=-\frac{3}{2}D(x-\frac{4R_{\nu}+5}{18(4R_{\nu}+15)}x^{3}), \nonumber \\
&& R_{\nu}=\frac{\rho^{(\nu})}{\rho^{(\nu)}+\rho^{(\gamma)}},\ x=\frac{k R}{H_{0}\sqrt{\Omega_{r}}},
\ D=arbitrary\ constant. \nonumber
\end{eqnarray}

In Table 1 one can find parameters used for the EC and the LCDM cosmology.

\begin{table}
\caption{Model parameters.}
\vspace{5mm}
\hspace{5mm} \begin{tabular}{| c || c | c | c | c | c | c | c |} \hline
model & $\Omega_{m}$ & $\Omega_{\Lambda}$ & h & $c_{0}$ & $z_{0};z_{1}$ & $\tau_{U}$ & $\Omega_{b}$ \\
\hline \hline
$\Lambda$CDM & 0.307 & 0.693 & 0.677  & -  & - & 13.83 Gyr & 0.045\\  \hline
EC  &  2  & 0  &  0.74  &  1.85  &  4;6 & 13.92 Gyr & 0.045   \\  \hline
\end{tabular}
\end{table}

We can now evaluate $\sigma_{8}(z)$ with the solutions of the coupled perturbation
equations (averaged with the top hat window function) \cite{Kolb} starting the integration
at the $R_{i}\equiv(1+z_{i})^{-1}=10^{-8}$:

\begin{eqnarray}
&&\sigma_{M}(R,z)=[<(\frac{\delta M}{M})^{2}(R,z)>]^{1/2},\
S_{8}=\sigma_{8}\sqrt{\frac{\Omega_{m}}{0.3}}, \nonumber \\
&&<(\frac{\delta M}{M})^{2}(R,z)>=N\int^{k_{max}}_{k_{min}}dk k^{2}
\delta_{(c)}^{2}(k,z)[\frac{\sin kR}{(kR)^{3}}-\frac{\cos kR}{(kR)^{2}}]^{2}, \nonumber \\
&&\sigma_{8}(z)\equiv \sigma_{M}(R=8 h^{-1}Mpc,z),\
\delta_{(c)}^{2}(k,z)=-\bar{\cal D}_{(c)}^{\mu}\bar{\cal D}_{(c)\mu}(k,z), \\
&& k_{min}=10^{-2}Mpc,\ k_{max}=10 Mpc. \nonumber
\end{eqnarray}

The reader can inspect our model of torsion in Fig. 1 and the age of the Universe as a function
of redshift in Fig. 2 for the EC and the LCDM cosmologies.
Our model of torsion is described by three parameters $c_{0}$, $z_{0}$ and $z_{1}$. The quantum spin
densities with the abundant right handed neutrinos \cite{Palle3} determine a torsion at very high redshifts
before the growth of structures: stars, black holes, clusters, globular clusters, galaxies and galaxy clusters.
We assume that at $z_{1}=6$ the right handed angular momentum of the Universe consisting of the nonlinear
structures begins to rise up to $z_{0}=4$. After redshift $z_{0}=4$ it scales according to angular momentum
of the nonlinear structures of the Zeldovich model \cite{Palle6}. Our model of torsion $Q(z)$ is crude and
nonanalytic function of redshift but it is properly normalized and describes well the essential features
of cosmography.

The redshift drifts for EC and LCDM models
are depicted in Fig. 3 using the well known formula \cite{Palle5,Liske}:
\begin{eqnarray*}
&&Q(R)=\bar{Q}(R)H_{0}, \\
&&\bar{Q}(R)=\sqrt{3}(1 + \frac{R-R_{0}}{R_{0}-R_{1}})[1-c_{0}+c_{0} R^{-3}]^{\frac{1}{2}},\
for\ R_{1} \leq R \leq R_{0}, \\
&&R=\frac{1}{1+z},\ R_{0}=\frac{1}{1+z_{0}},\ R_{1}=\frac{1}{1+z_{1}}, \\
&&\bar{Q}(R)=\sqrt{3}[1-c_{0}+c_{0} R^{-3}]^{\frac{1}{2}},\ for\ R_{0} \leq R \leq 1, \\
&&\Omega_{Q}= -1\ \Longleftrightarrow \ \bar{Q}(R=1)=\sqrt{3},\ \
\bar{Q}(R \leq R_{1}) = 0 ,
\end{eqnarray*}
\begin{eqnarray*}
&&LCDM: \tau_{U} = \frac{9.7776}{h}\int^{1}_{10^{-8}}\frac{dR}{R}
[\Omega_{r}R^{-4}+\Omega_{m}R^{-3}+\Omega_{\Lambda}]^{-\frac{1}{2}}Gyr,  \\
&& EC: \tau_{U} = \frac{9.7776}{h}\int^{1}_{10^{-8}}\frac{dR}{R}
[\Omega_{r}R^{-4}+\Omega_{m}R^{-3}-\frac{1}{3}\bar{Q}^{2}(R)]^{-\frac{1}{2}}Gyr, \\
&& redshift\ drift\equiv\frac{\Delta z}{\Delta t}=(1+z)H_{0}-H(z).
\end{eqnarray*}

\begin{figure}[htb]
\centerline{
\includegraphics[width=12cm]{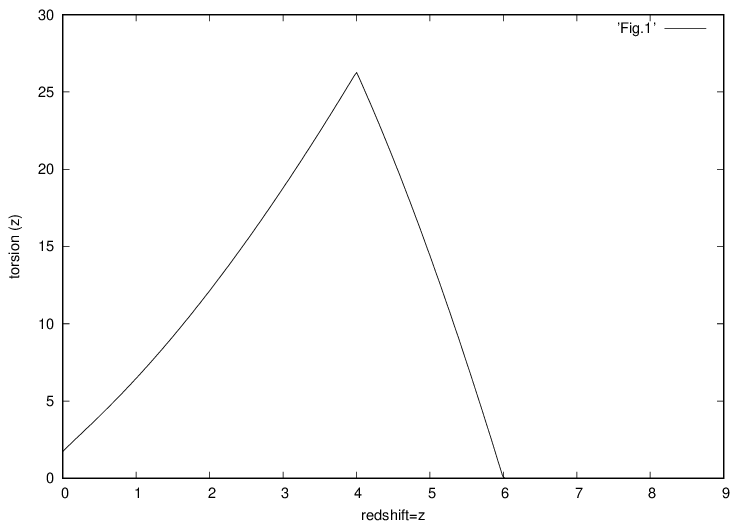}}
\caption{Model of torsion $\bar{Q}$(z) used in numerical evaluations.}
\end{figure}

One can observe large discrepancies between the EC and LCDM cosmologies at high redshifts for the
age of the Universe and redshift drifts.

\begin{figure}[htb]
\centerline{
\includegraphics[width=12cm]{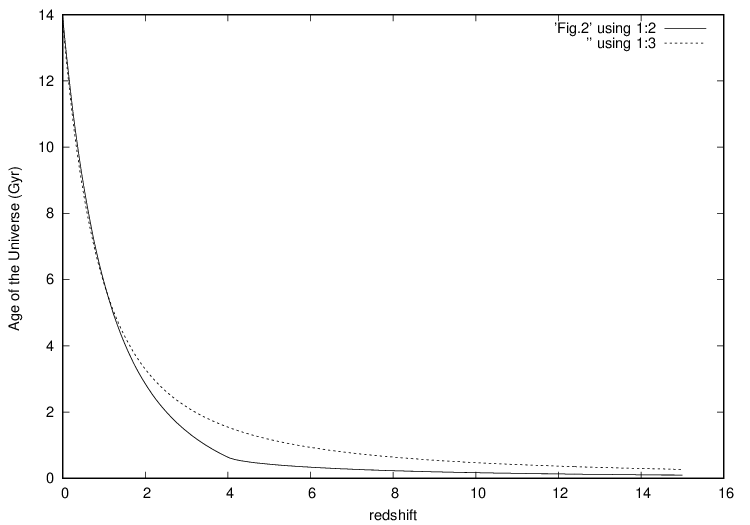}}
\caption{Age of the Universe for the EC (solid line) and LCDM (dotted line) models.}
\end{figure}

\begin{figure}[htb]
\centerline{
\includegraphics[width=12cm]{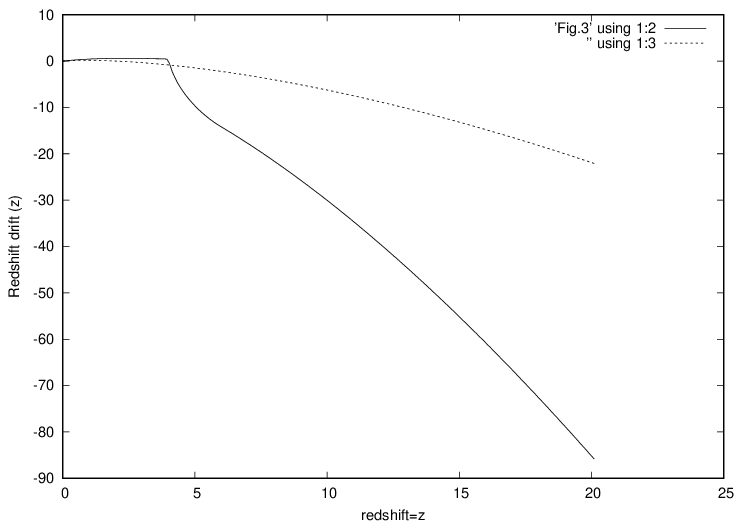}}
\caption{Redshift drifts for the EC (solid line) and LCDM (dotted line) models in the unit
$100 km s^{-1} Mpc^{-1}$.}
\end{figure}

The main result of this paper is the redshift dependence of the $\sigma_{8}(z)$ shown in Fig. 4
normalized to arbitrary $\sigma_{8}(0)$ for EC and LCDM evaluated within the same approximations.

It is not a surprise to observe substantially enhanced EC $\sigma_{8}(z)$ at high redshift
with respect to low redshift in contrast to LCDM $\sigma_{8}(z)$ \cite{DESI,Euclid,DES}.
We study the evolution of $\sigma_{8}(z)$ with the acceleration $\lambda \neq 0$ and find
negligible influence for $\lambda=10^{-3}=const.$ or for $\lambda (z)$ (Fig. 5) \cite{Palle12}.

\begin{figure}[htb]
\centerline{
\includegraphics[width=12cm]{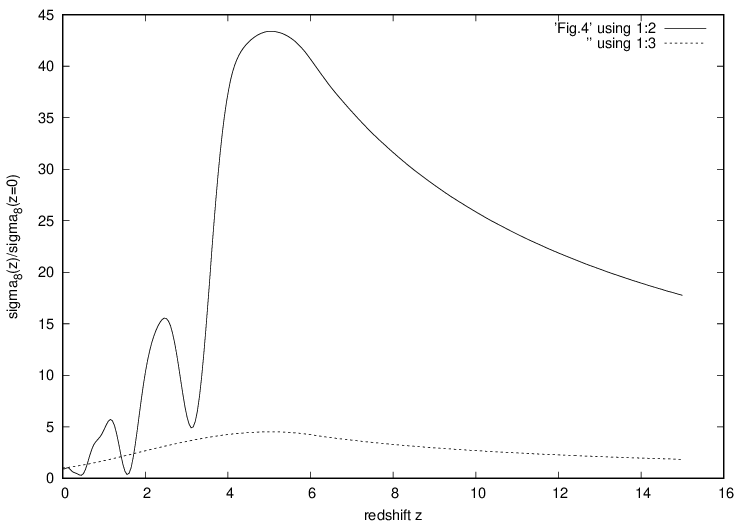}}
\caption{$\sigma_{8}(z)/\sigma_{8}(z=0)$ for the EC (solid line) and LCDM (dotted line) models.}
\end{figure}

\begin{figure}[htb]
\centerline{
\includegraphics[width=12cm]{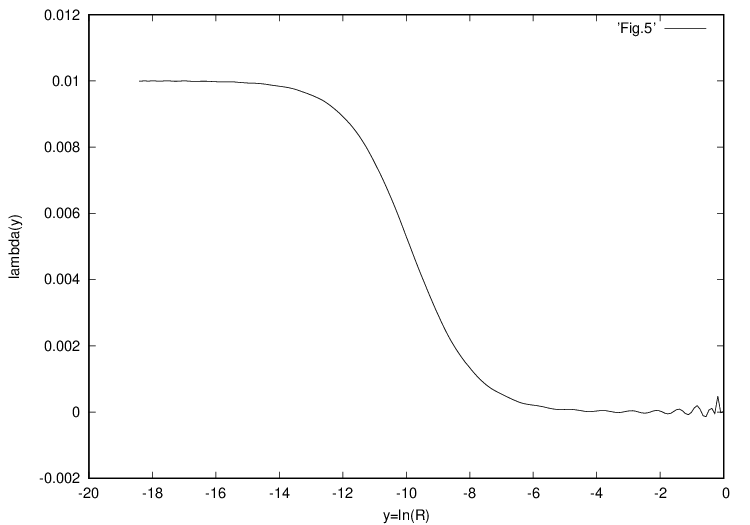}}
\caption{$\lambda$(y) for initial $\lambda(R=10^{-8})=10^{-2}$.}
\end{figure}

Thus, the growth of structures is stronger and turns up earlier and faster in the evolution in the EC than in the LCDM model
due to much larger mass density and $\sigma_{8}(z)$ of the EC model at high redshifts \cite{ALMA,JWST1,JWST2,JWST3}.

Let us mention two additional problems for the LCDM model: the Hubble tension and
the large scale anomaly of the CMB, namely
the most precise TT CMB anisotropy spectrum measured by Planck satellite at low mutipoles (large scales) is
not compatible with large multipoles (small scales).

The future theoretical study of the EC model requests inevitably the N-body numerical simulations
with the EC equations of motion and evaluation of the angular momentum of the Universe at every moment
of the evolution. This is the only method available to find the evolution of torsion $Q(z)$ as a function of
the chiral asymmetry and the magnitude of the vorticity.

\vspace{10mm}
{\bf Appendix A}
\vspace{5mm}
\newline
A decomposition of the conformal tensor into electric and magnetic parts is a quite often procedure.
Let us start with the definition of the Weyl tensor in Riemannian spacetime:
\begin{eqnarray*}
C_{\sigma\lambda\mu\nu}&\equiv& R_{\sigma\lambda\mu\nu}+\frac{1}{2}(g_{\sigma\nu}R_{\lambda\mu}
+g_{\lambda\mu}R_{\sigma\nu}-g_{\sigma\mu}R_{\lambda\nu}-g_{\lambda\nu}R_{\sigma\mu}) \\
&&+\frac{1}{6}(g_{\sigma\mu}g_{\lambda\nu}-g_{\sigma\nu}g_{\lambda\mu})R.
\end{eqnarray*}

The standard definitions of the electric and magnetic parts are \cite{Ellis4}:
\begin{eqnarray*}
E_{\mu\nu}\equiv u^{\lambda}u^{\kappa} C_{\mu\lambda\nu\kappa},\
B_{\mu\nu}\equiv \frac{1}{2}u^{\lambda}u^{\kappa}\eta_{\mu\lambda}^{\ \ \rho\alpha}C_{\kappa\nu\rho\alpha},\
\eta_{0123}=-[-det(g_{\mu\nu})]^{1/2}.
\end{eqnarray*}

Decomposition of the Weyl tensor follows as:
\begin{eqnarray*}
C^{\rho\mu}_{(E)\ \ \nu\lambda}=4(u^{[\rho}u_{[\nu}E^{\mu]}_{\ \lambda}-h^{[\rho}_{[\nu} E^{\mu]}_{\ \lambda]}),
\ C^{(B)}_{\rho\mu\nu\lambda}=2(\eta_{\rho\mu\sigma}u_{[\nu}B_{\lambda]}^{\ \sigma}
+\eta_{\nu\lambda\sigma}u_{[\rho}B_{\mu]}^{\ \sigma}).
\end{eqnarray*}

Direct computation with the metric in Eq.(3) gives:
\begin{eqnarray*}
C_{\sigma\lambda\mu\nu}-C_{(E)\sigma\lambda\mu\nu}-C_{(B)\sigma\lambda\mu\nu}
={\cal O}_{\sigma\lambda\mu\nu}(\frac{d \lambda}{d t}\omega).
\end{eqnarray*}

Thus, if $\frac{d \lambda}{d t}\omega \neq 0$ (vorticity and time derivative of $\lambda$ do not vanish),
few components of the tensor of difference do not vanish and the
standard decomposition of the Weyl tensor is not valid.

\vspace{10mm}
{\bf Appendix B}
\vspace{5mm}
\newline
We derive in this appendix the evolution equations for gauge invariant density contrasts in the radiation and
matter dominated eras without vorticity, accelerations and torsion, but assuming the well known
form of the evolution derived from the gauge dependent formalism. This is the way how one can fix
the form of the gauge invariant quantities.

Let us assume the following form of the gauge invariant quantities in the radiation dominated era
($p=\frac{1}{3}\rho$):
\begin{eqnarray*}
{\cal D}_{\mu}\equiv R^{\alpha}\rho^{-1\ (3)}\nabla_{\mu} \rho,\
Z_{\mu}\equiv R^{\alpha\ (3)}\nabla_{\mu} \theta. \\
\end{eqnarray*}

The standard procedure by taking into account field equations and identities leads to
equations:
\begin{eqnarray*}
\dot{{\cal D}_{\mu}}&=&\frac{\Theta}{3}(\alpha -1){\cal D}_{\mu}-\frac{4}{3}Z_{\mu}, \\
\dot{Z_{\mu}}&=& \frac{\Theta}{3}(\alpha -3)Z_{\mu}-\frac{1}{3}\Theta^{2}{\cal D}_{\mu}.
\end{eqnarray*}

We include the known form for the growing solution of the density contrast $\delta \equiv
(-{\cal D}_{\mu}{\cal D}^{\mu})^{1/2} \propto R^{2}$ into the above equations and
find the algebraic equation for $\alpha$:
\begin{eqnarray*}
\alpha^{2} -6 \alpha + 5 = 0 \Rightarrow \alpha_{1}=1,\ \alpha_{2}=5 .
\end{eqnarray*}

Repeating the same procedure in the matter dominated era ($p=0$) with definitions:
\begin{eqnarray*}
{\cal D}_{\mu}\equiv R^{\beta}\rho^{-1\ (3)}\nabla_{\mu} \rho,\
Z_{\mu}\equiv R^{\beta\ (3)}\nabla_{\mu} \theta. \\
\end{eqnarray*}

we get the equations:
\begin{eqnarray*}
\dot{{\cal D}_{\mu}}&=&\frac{\Theta}{3}(\beta -1){\cal D}_{\mu}-Z_{\mu}, \\
\dot{Z_{\mu}}&=& \frac{\Theta}{3}(\beta -3)Z_{\mu}-\frac{1}{6}\Theta^{2}{\cal D}_{\mu}.
\end{eqnarray*}

Density contrast grows as $\delta \propto R$ and the equation for $\beta$ is:
\begin{eqnarray*}
2 \beta^{2} -9 \beta + 7 = 0 \Rightarrow \beta_{1}=1,\ \beta_{2}=\frac{7}{2} .
\end{eqnarray*}

It follows that the most natural form for a density contrast in both radiation and
matter dominated eras is $\alpha_{1}=\beta_{1}=1$.

\end{document}